\documentclass[journal]{IEEEtran}

\usepackage{amssymb}
\usepackage[cmex10]{amsmath}
\usepackage{stfloats}
\usepackage{graphicx}
\usepackage{subfigure}
\usepackage{tabularx}
\usepackage{epsfig,epsf,color,balance,cite}
\usepackage{verbatim}
\usepackage{url}
\usepackage{bm}

\definecolor{myc1}{rgb}{0,0,0}
\usepackage{cases}

\usepackage{algorithm}
\usepackage{algorithmic}
\begin{document}

\title{Joint Altitude, Beamwidth, Location and Bandwidth Optimization for UAV-Enabled  Communications}

\author{
\IEEEauthorblockN{Zhaohui Yang,
                  Cunhua Pan,
                  Mohammad Shikh-Bahaei,
                  Wei Xu,    
                  Ming Chen,
                  Maged Elkashlan, and Arumugam Nallanathan
                  }
                  \vspace{-3em}
\thanks{Z. Yang and M. Shikh-Bahaei are with Centre for Telecommunications Research, King's College London, London, UK (Emails: \{yang.zhaohui, m.sbahaei\}@kcl.ac.uk)
 W. Xu and M. Chen are with the National Mobile Communications Research
Laboratory, Southeast University, Nanjing 210096, China  (Email: \{wxu, chenming\}@seu.edu.cn).}
 \thanks{C. Pan M. Elkashlan, and A. Nallanathan are with the School of Electronic Engineering and Computer Science, Queen Mary, University of London, London E1 4NS, U.K. (Emails: \{c.pan,  maged.elkashlan, a.nallanathan\}@qmul.ac.uk).}
}
\maketitle
\begin{abstract}
This letter investigates an uplink power control problem for unmanned aerial vehicles (UAVs) assisted wireless communications.
We jointly optimize the UAV's flying altitude, antenna beamwidth, UAV's location and ground terminals' allocated bandwidth and transmit power to minimize the sum uplink power subject to the minimal rate demand.
An iterative algorithm is proposed with low complexity to obtain a suboptimal solution.
Numerical results show that the proposed algorithm can achieve good performance in terms of uplink sum power saving.
\end{abstract}


\IEEEpeerreviewmaketitle

 \vspace{-1em}

\section{Introduction}
Unmanned aerial vehicles (UAVs) assisted wireless communications have attracted considerable attention recently due to its maneuverability and increasing affordability \cite{Zeng2017Mag}.
Compared to conventional wireless communications,
UAV-enabled wireless communications can provide higher wireless connectivity in areas without infrastructure coverage and achieve higher capacity for line-of-sight (LoS) communication links with the ground terminals (GTs).

To fully exploit the design degrees of freedom for UAV-enabled communications,
it is crucial to investigate the UAV mobility in the three-dimensional space.
{\color{myc1}{In \cite{6863654}, the altitude of UAV was optimized to provide maximum radio coverage on the ground.
For an underlaid device-to-device (D2D) communication network with one UAV,
the optimal values for the UAV altitude were analyzed in \cite{7412759} for the maximum system sum rate and coverage probability.
Considering the adjustable UAVs' locations over time, the UAV number and trajectory optimization problems were respectively considered in \cite{7762053} and \cite{7888557}.
Further optimizing user-UAV association, \cite{7875131} investigated the sum power minimization problem of the UAV.
Different from \cite{6863654,7412759,7762053,7888557,7875131} with fixed-beamwidth antenna, the beamwidth of the directional antenna and the altitude of the UAV were jointly optimized in \cite{He2017CL} to improve the system throughput.
However, the optimal beamwidth was only examined numerically and simple equal bandwidth allocation was assumed in \cite{He2017CL}, even though proper bandwidth allocation can further enhance the system performance.}}

In this letter, we aim to minimize the sum power for an uplink UAV-enabled wireless communication. 
{\color{myc1}{There are two main contributions.
One contribution is that we consider joint altitude, beamwidth, location and bandwidth allocation, and an algorithm is proposed by solving three subproblems iteratively, where each subproblem can be solved optimally.
We also provide the complexity analysis of the proposed algorithm. Numerical results verify that the proposed algorithm outperforms the existing algorithms with fixed beamwidth or bandwidth allocation in terms of sum power, especially when the minimal rate demand is high.
The other contribution is to effectively obtain the optimal beamwidth with the bisection method when the pathloss exponent is two, and to obtain the optimal solution in closed form for bandwidth allocation subproblem.}}

 \vspace{-1em}

\section{System Model and Problem Formulation}

Consider an uplink UAV-enabled wireless communication system with one flying UAV and $K$ GTs. 
The UAV is deployed as a flying BS with horizontal and vertical location $\pmb y=(y(1),y(2))$ at hight $H$.
The  horizontal and vertical location of  GT $k$ is denoted by $\pmb x_k =(x_k(1), x_k (2))$, and the hight of each GT is assumed to be zero compared with the hight of the UAV.


Assume that the UAV is equipped with a directional antenna with adjustable beamwidth, while each GT is equipped with an omnidirectional antenna with unit gain.
The azimuth and elevation half-power beamwidths of the UAV antenna are equal, which are both denoted by $2\Theta\in(0, \pi)$.
According to \cite[Eq.~(2-51)]{constantine2016antenna}, the antenna gain in the direction with azimuth angle $\theta$ and elevation angle $\phi$ can be modeled as
\begin{equation}\label{syseq1}
\begin{aligned}
&G=
  \begin{cases}
\frac{G_0}{\Theta^2} &\!\! \mbox{if $0\leq \theta\leq \Theta$ and $0\leq \phi\leq \Theta$} \\
 g \approx 0 &\!\!  \mbox{otherwise},
  \end{cases}
\end{aligned}
\end{equation}
where $G_0\approx2.2846$,
and {\color{myc1}{$g$ means the channel gain outside the beamwidth of the antenna}}.
For simplify, we set $g=0$.
We consider the case that the GTs are located outdoors, and the channel between each GT and the UAV is mainly a LoS path.
The uplink channel gain between GT $k$ and the UAV is
\begin{equation}\label{syseq2}
g_k= \frac {g_0} { \left( \| \pmb y- \pmb x_k \|^2 +H^2 \right)^{\frac \alpha 2} },
\end{equation}
where $\|\cdot\|$ denotes the Euclidian norm,
$g_0$ is the channel power gain at the reference distance 1 m,
$H$ is the hight of the UAV,
$ \left( \|\pmb y -\pmb x_k\|^2 + H^2 \right) ^{\frac 12 } $ is the distance between GT $k$ and the UAV,
and $\alpha\geq 2$ is the pathloss exponent.
Based on (\ref{syseq1}) and (\ref{syseq2}), the uplink achievable rate of GT $k$ in the coverage area of the UAV is
\begin{equation}\label{syseq3}
r_k =w_k \log_2 \left( 1+ \frac{ p_k g_0 G_0}
{w_k\sigma^2 \Theta^2{ \left( \| \pmb y- \pmb x_k \|^2 +H^2 \right)^{\frac \alpha 2} }}
\right),
\end{equation}
where $w_k$ is the allocated bandwidth for GT $k$,
$p_k$ is the transmit power of GT $k$,
$\sigma^2$ is the noise power density and $w_k \sigma^2$ is the noise power for decoding the information of GT $k$ at the UAV side.
For GT $k$, the minimal rate constraint $r_k\geq R_k$ should be satisfied.
Since we aim at minimizing uplink sum power of all GTs, it is always energy saving to transmit with minimal rate.
Setting $r_k=R_k$ in (\ref{syseq3})
, we have
\begin{equation}\label{PAet1}
p_k= a {w_k\left( 2^{ \frac {R_k} {w_k} } -1 \right){ \Theta^2{ \left( \| \pmb y- \pmb x_k \|^2 +H^2 \right)^{\frac \alpha 2} }}} ,
\end{equation}
where $a =\frac {\sigma^2} {g_0G_0}$.

{\color{myc1}{We aim at minimizing the uplink sum power of all GTs whilst satisfying the minimal rate constraints}}.
Mathematically, the sum power minimization problem is formulated as
\vspace{-1em}
\begin{subequations}\label{PAmin2}
\begin{align}
\mathop{\min}_{  H, \Theta, \pmb y,\pmb w}\:\;
& \sum_{k=1}^{ K} a {w_k\left( 2^{ \frac {R_k} {w_k} } -1 \right){ \Theta^2{ \left( \| \pmb y- \pmb x_k \|^2 +H^2 \right)^{\frac \alpha 2} }}}\\
\textrm{s.t.}\qquad \!\!\!
&a {w_k\left( 2^{ \frac {R_k} {w_k} } -1 \right){ \Theta^2{ \left( \| \pmb y- \pmb x_k \|^2 +H^2 \right)^{\frac \alpha 2} }}}\leq P_k,
\nonumber\\
&\qquad\qquad\qquad\qquad\qquad\qquad \forall k=1, \cdots, K\\
& \| \pmb y- \pmb x_k \|^2 \leq H^2 \tan^2 \Theta, \quad \forall k=1, \cdots, K\\
& \sum_{k=1}^K w_k \leq B \\
&H_{\min} \leq H \leq H_{\max}, \Theta_{\min}\leq \Theta \leq \Theta_{\max}\\
&  w_k \geq0, \quad \forall k=1, \cdots, K.
\end{align}
\end{subequations}
where $\pmb w=(w_1, \cdots, w_K)$,
$B$ is the  maximal bandwidth of the system,
$P_k$ is the maximum transmit power of GT $k$,
$[H_{\min}, H_{\max}]$ is the feasible region of height $H$ determined by obstacle heights and authority regulations,
and $[\Theta_{\min}, \Theta_{\max}]$ is the feasible region of half-beamwidth determined by practical antenna beamwidth tuning technique.
Constraints in (\ref{PAmin2}c) ensure that all GTs are in the coverage area of the UAV.

\vspace{-1em}
\section{Proposed Algorithm}
\vspace{-0.5em}

Due to nonconvex objective function (\ref{PAmin2}a) and constraints (\ref{PAmin2}b)-(\ref{PAmin2}c), Problem (\ref{PAmin2}) is a nonconvex problem.
It is generally hard to obtain the globally optimal solution to Problem (\ref{PAmin2}).
To solve this problem, we propose an iterative algorithm with low complexity through sequently optimizing $(H,\Theta)$, $\pmb y$ and $\pmb w$.
It is fortunate that we can globally optimize each variable with other variables fixed in each step.

\vspace{-1.25em}
\subsection{Optimal Altitude and Beamwidth}
\vspace{-0.25em}
{\color{myc1}{With fixed $\pmb y$ and $\pmb w$,
Problem (\ref{PAmin2}) is formulated as}}
\begin{subequations}\label{PAOABmin1}
\begin{align}
\mathop{\min}_{  H, \Theta}\quad\:\;
& \sum_{k=1}^{ K} A_k{ \Theta^2{ ( D_k +H^2  )^{\frac \alpha 2} }}\\
\textrm{s.t.}\qquad \!\!
&A_k{ \Theta^2{ (D_k +H^2)^{\frac \alpha 2} }} \leq P_k, \quad\forall k=1, \cdots, K\\
&   H^2 \tan^2 \Theta \geq D_{\max}\\
&H_{\min} \leq H \leq H_{\max}, \Theta_{\min}\leq \Theta \leq \Theta_{\max},
\end{align}
\end{subequations}
where $A_k=a w_k\left( 2^{ \frac {R_k} {w_k} } -1 \right)$,
$D_k= \| \pmb y- \pmb x_k \|^2$, and $D_{\max}=\max_{k=1,\cdots, K} D_k$.
{\color{myc1}{Denoting $H^*$ as optimal value of $H$ and observing that the objective function (\ref{PAOABmin1}a) is an increasing function in $H$ with given $\Theta$,  we can claim that
\begin{equation}\label{PAOABeq1}
H^*=\max\left\{\frac{\sqrt{D_{\max} }}{ \tan\Theta }, H_{\min} \right\},
\end{equation}
for the optimal solution.
This claim can be proved by the contradiction method.
If $(H, \Theta)$ is the optimal solution of Problem (\ref{PAOABmin1}) with $H>H^*$, we find that solution $(H^*, \Theta)$ is also a feasible solution of Problem (\ref{PAOABmin1}) with $\sum_{k=1}^{ K} A_k{ \Theta^2{ ( D_k +(H^*)^2  )^{\frac \alpha 2} }}<\sum_{k=1}^{ K} A_k{ \Theta^2{ ( D_k +H^2  )^{\frac \alpha 2} }}$, which contradicts the hypothesis that $(H, \Theta)$ is the optimal solution.}}
Based on (\ref{PAOABeq1}), we consider the value of $H^*$ in the following two cases.

1) Case 1: With $H^*=H_{\min}$, Problem (\ref{PAOABmin1}) is equivalent to
\begin{subequations}\label{PAOABmin2}
\begin{align}
\mathop{\min}_{ \Theta}\quad\:\;
&   \Theta  \\
\textrm{s.t.}\qquad \!\!
&A_k{ \Theta^2{ (D_k +H_{\min}^2)^{\frac \alpha 2} }} \leq P_k, \quad\forall k=1, \cdots, K\\
&  H_{\min}^2 \tan^2 \Theta \geq D_{\max}\\
&  \Theta_{\min}\leq \Theta \leq \Theta_{\max}.
\end{align}
\end{subequations}
{\color{myc1}{Since Problem (\ref{PAOABmin2}) is a minimization of $\Theta$, the optimal solution  is thus
\begin{equation}\label{PAOABeq2}
\Theta^*=\max\left\{\arctan\frac{\sqrt{ D_{\max} }}{ H_{\min} }, \Theta_{\min} \right\},
\end{equation}
which is the minimal value of $\Theta$ satisfying (\ref{PAOABmin2}b), (\ref{PAOABmin2}c) and (\ref{PAOABmin2}d).}}
Note that Problem (\ref{PAOABmin2}) is feasible if and only if
\begin{equation}\label{PAOABeq2}
\Theta^* \leq \min\left\{\min_{k=1,\cdots, K} \sqrt{\frac{P_k}{A_k { (D_k +H_{\min}^2)^{\frac \alpha 2} }}} , \Theta_{\max}  \right\}.
\end{equation}

2) Case 2: With $H^*=\frac{\sqrt{D_{\max}}}{\tan\Theta}$, Problem (\ref{PAOABmin1}) becomes
\begin{subequations}\label{PAOABmin3}
\begin{align}
\mathop{\min}_{ \Theta}\quad\:\;
& \sum_{k=1}^{ K} A_k{ \Theta^2{ \left(  D_k +\frac {D_{\max}} {\tan^2 \Theta}   \right)^{\frac \alpha 2} }}\\
\textrm{s.t.}\qquad \!\!
&A_k{ \Theta^2{ \left( D_k + \frac{D_{\max}} {\tan^2 \Theta}   \right)^{\frac \alpha 2} }} \leq P_k, \quad\forall k=1, \cdots, K\\
&  H_{\min}^2 \tan^2 \Theta \leq D_{\max}\\
& \Theta_{\min}\leq \Theta \leq \Theta_{\max}.
\end{align}
\end{subequations}
Due to the complicated objective function (\ref{PAOABmin3}a), it is generally difficult to obtain the optimal $\Theta^*$ of Problem (\ref{PAOABmin3}) in closed form.
{\color{myc1}{$\Theta^*$ can be obtained via a one-dimension exhaustive search over $[\Theta_{\min}, \Theta_{\max}]$.}}

Specifically, for the special case where pathloss exponent $\alpha=2$, we can fortunately obtain the optimal $\Theta^*$ through a simple bisection method.
{\color{myc1}{When the GTs are located outdoors in rural areas, and the communication channel between the UAV and each GT is dominated by the LoS path, i.e., $\alpha=2$ \cite[Eq. (2)]{He2017CL}.}}
For $\alpha=2$, we define function
\begin{equation}
f_k(x)={ x^2 { \left( D_k +\frac {D_{\max}} {\tan^2 x}   \right)}}, \quad   x\in[0,\pi/2),
\end{equation}
\begin{equation}
h_1(x)=(\cot x -x  -  x\cot ^2 x)\cot x,
\end{equation}
\begin{equation}
h_2(x)=x + 2 x \cos^2(x) - 3/2 \sin(2 x),
\end{equation}
and then we have
\begin{equation}\label{PAOABeq2_1}
f_k'(x)=2x\left(
D_k + D_{\max} h_1(x)
\right),
\end{equation}
\begin{equation}\label{PAOABeq2_2}
h_1'(x) = \csc^4(x)  h_2(x),
\end{equation}
\begin{equation}\label{PAOABeq2_2}
h_2'(x) =-4 (x \cos(x) - \sin(x)) \sin(x),
\end{equation}
for $x\in[0,\pi/2)$.
Since $x\leq\tan x$ for $x\in[0,\pi/2)$, we have $x \cos(x) - \sin(x)\leq0$, i.e., $h_2'(x)\geq0$.
As a result,  $h_2(x)\geq h_2(0)=0$, $h_1'(x)=\csc^4xh_2(x)\geq 0$, i.e., $h_1(x)$ is an increasing function, and $-2/3 =\lim_{x\rightarrow 0+} h_1(x) \leq h_1(x) \leq \lim_{x\rightarrow (\pi/2)-} h_1 (x)=0$.
Due to that $x\geq 0$, $f'_k(x)\geq 0$ is equivalent to $D_k+D_{\max}h_1(x)\geq 0$.
To show the monotonicity of $f_k(x)$, we consider the following two situations:
\begin{itemize}
  \item If $D_k-\frac 2 3D_{\max}\geq 0$, then $D_k+D_{\max}h_1(x)\geq 0$ for all $x\in[0, \pi/2)$, i.e., $f_k(x)$ is monotonically increasing.
  \item If $D_k-\frac 2 3D_{\max}< 0$, there must exist one solution $x_k$ such that $D_k+D_{\max}h_1(x_k) = 0$ due to the fact that $D_k \geq 0$ and $h_1(x)$ is a continuous function.
      In this situation, $f_k(x)$ first decreases for $x\in[0,x_k]$ and then increases with $x\in(x_k,\pi/2)$.
\end{itemize}
According to the above analysis, $f_k(\Theta)\leq P_k$ is equivalent to $\Theta_k^{\min}\leq \Theta\leq \Theta_k^{\max}$, where $\Theta_k^{\min}$
and $\Theta_k^{\max}$ can be obtained by using the bisection method.
As a result, constraints (\ref{PAOABmin3}b)-(\ref{PAOABmin3}d) can be equivalently transformed to
\begin{equation}\label{PAOABeq2_3}
\bar\Theta_{\min}\leq \Theta\leq \bar\Theta_{\max},
\end{equation}
where $\bar\Theta_{\min}=\max\{\max_{k=1,\cdots,K}\Theta_k^{\min}, \Theta_{\min}\}$,
$\bar\Theta_{\max}=\min\{\min_{k=1,\cdots,K}\Theta_k^{\max}, \Theta_{\max}, \arctan (\sqrt{D_{\max}}/H_{\min})\}$.

Based on the definition of $f_k(x)$, the objective function (\ref{PAOABmin3}a) can be expressed as $f(\Theta)=\sum_{k=1}^Kf_k(\Theta)$.
We have $f'(\Theta)=\sum_{k=1}^Kf_k'(\Theta)=2x(\sum_{k=1}^KD_k + K D_{\max} h_1(\Theta))$.
To show the monotonicity of $\Theta$ in $[\bar\Theta_{\min}, \bar \Theta_{\max}]$, we also consider the following three situations:
\begin{itemize}
  \item If $\sum_{k=1}^KD_k + KD_{\max}h_1(\bar\Theta_{\min})\geq 0$, then $\sum_{k=1}^KD_k + KD_{\max}h_1(\Theta)\geq 0$ for $\Theta\in[\bar\Theta_{\min}, \bar \Theta_{\max}]$, i.e., $f(\Theta)$ is monotonically increasing.
      The optimal beamwidth is $\Theta^*=\bar\Theta_{\min}$.
  \item If 
  $\sum_{k=1}^KD_k + KD_{\max}h_1(\bar\Theta_{\max})< 0$, $f(\Theta)$ is monotonically decreasing and $\Theta^*=\bar \Theta_{\max}$.
  \item If $\sum_{k=1}^KD_k + KD_{\max}h_1(\bar\Theta_{\min})< 0$ and $\sum_{k=1}^KD_k + KD_{\max}h_1(\bar\Theta_{\max})> 0$, there must exist one solution $\bar\Theta$ such that $\sum_{k=1}^KD_k + KD_{\max}h_1(\Theta)= 0$.
      In this situation, $f(\bar\Theta)$ first decreases for $x\in[\bar\Theta_{\min},\bar\Theta]$ and then increases with $x\in(\bar\Theta,\bar\Theta_{\max}]$, i.e., $\Theta^*=\bar\Theta$.
\end{itemize}

Note that Problem (\ref{PAOABmin3}) is feasible if and only if $\bar\Theta_{\min}\leq \bar\Theta_{\max}$.
By comparing the objective values of the solutions obtained in the above two cases, the one with lower objective value is chosen as the optimal solution to Problem (\ref{PAOABmin1}).

 \vspace{-1em}
\subsection{Optimal Location Planning}

For Problem (\ref{PAmin2}) with fixed $(H, \Theta)$ and $\pmb w$,
the location planning problem can be formulated as
\begin{subequations}\label{PALPmin1}
\begin{align}
 \!\!\!\mathop{\min}_{ \pmb y} \:\;
& \sum_{k=1}^{ K} C_k { \left( \| \pmb y- \pmb x_k \|^2 +H^2 \right)^{\frac \alpha 2} }\\
\textrm{s.t.} \:\;
&   {  \| \pmb y- \pmb x_k \|^2 +H^2  } \leq \bar E_k,
\quad \forall k=1, \cdots, K\\
& \| \pmb y- \pmb x_k \|^2 \leq H^2 \tan^2 \Theta, \quad \forall k=1, \cdots, K,
\end{align}
\end{subequations}
where $C_k=a  w_k\left( 2^{ \frac {R_k} {w_k} } -1 \right)\Theta^2$,
and $\bar E_k=\left(\frac{P_k}{C_k}\right)^{\frac 2\alpha}$.
Since $\|\pmb y-\pmb x_k\|^2$ is a convex function and $x^{\frac{\alpha}{2}}$ is convex and nondecreasing, $\left( \| \pmb y- \pmb x_k \|^2 +H^2 \right)^{\frac \alpha 2}$ is convex based on the scalar composition property of convex functions \cite{boyd2004convex}.
As a result, Problem (\ref{PALPmin1}) is a convex problem, {\color{myc1}{which can be effectively solved via the standard interior point method}}.

 \vspace{-1em}
\subsection{Optimal Bandwidth Allocation}
It remains to investigate the bandwidth allocation with fixed location, altitude and beamwidth.
Define function $u_k(x)=x 2^{\frac {R_k} x}-x$ for $x> 0$, and we have
\begin{equation}\label{PABandeq1}
u_k'(x)=2^{\frac {R_k} x}-\frac{(\ln 2)R_k2^{\frac {R_k} x}}{x}-1,
u_k''(x)= \frac{(\ln2)^2R_k^22^{\frac {R_k} x}}{x^3}> 0.
\end{equation}
From (\ref{PABandeq1}), we observe that function $u_k(x)$ is a convex function, which indicates that Problem (\ref{PAmin2}) is a convex problem with fixed $(H, \Theta)$ and $\pmb y$.
Based on (\ref{PABandeq1}),
we have $u_k'(x)<\lim_{x\rightarrow+\infty}u_k'(x)=0$, i.e., $u_k(x)$ is a monotonically decreasing function, which is helpful in transforming constraints (\ref{PAmin2}b).
With optimized $(H, \Theta)$ and $\pmb y$, Problem (\ref{PAmin2}) is equivalent to
\begin{subequations}\label{PABand}
\begin{align}
\mathop{\min}_{  \pmb w }\qquad
& \sum_{k=1}^{ K}F_k {w_k\left( 2^{ \frac {R_k} {w_k} } -1 \right)}\\
\textrm{s.t.}\qquad \:\;\!\!
& \sum_{k=1}^K w_k \leq B \\
&  w_k \geq W_k \quad \forall k=1, \cdots, K,
\end{align}
\end{subequations}
where $F_k=a { \Theta^2{ \left( \| \pmb y- \pmb x_k \|^2 +H^2 \right)^{\frac \alpha 2} }}$, $W_k=u_k^{-1}\left(\frac{P_k}
{b_k} \right)$, and $u_k^{-1}(x)$ is the inverse function of $u_k(x)$.
{\color{myc1}{
The lagrangian of convex Problem (\ref{PABand}) is
\begin{equation}
\mathcal L(\pmb{ w}, \lambda)=
\sum_{k=1}^{ K}F_k {w_k\left( 2^{ \frac {R_k} {w_k} } -1 \right)} + \lambda \left(\sum_{k=1}^K w_k - B \right),
\end{equation}
where $\lambda$ is the non-negative Lagrange multiplier associated with constraint (\ref{PABand}b).
According to \cite{boyd2004convex} and \cite[Appendix~A]{Yang2017TVT}, the KKT conditions of (\ref{PABand}) are
\begin{equation}\label{re1KKT1}
\frac{\partial \mathcal L}{\partial w_k}=
F_{k} \left(\textrm{2}^{\frac{R_{k}} {w_{k}}}-\frac{(\ln2)R_{k}} {w_{k}} \textrm 2^{\frac{R_{k}} {w_{k}}} -1\right) + \lambda
=0
\end{equation}
From (\ref{re1KKT1}), we have
\begin{equation}\label{re1eqapp1}
\lambda=F_{k} \left(-\textrm{e}^{\frac{(\ln 2)R_{k}} {w_{k}}}+\frac{(\ln2)R_{k}} {w_{k}} \textrm e^{\frac{(\ln 2)R_{k}} {w_{k}}} +1\right) .
\end{equation}
Define function
$u(x)= x \textrm e^x -\textrm e^x +1$, $x \geq 0$.
We have $u'(x)=x \textrm e^x >0$, $\forall x >0$.
Thus, function $u(x)$ is strictly increasing and $u(x) > u(0) =0$, $\forall x >0$.
Based on (\ref{re1eqapp1}) and (\ref{PABand}c), we have
\begin{equation}\label{re1eqapp1_3}
{w_{k}}=\max\left\{\frac {(\ln 2)R_{k}}{u^{-1}\left({\frac{\lambda} {F_{k}}}\right)}, {W_k} \right\},\quad \forall k=1,\cdots, K,
\end{equation}
where $u^{-1}(x)$ is the inverse function of $u(x)$.
According to (\ref{re1eqapp1}), $\lambda=F_k u\left({\frac{(\ln 2)R_{k}} {w_{k}}}\right)>0$, which implies that (\ref{PABand}b) holds with equality.
Plugging (\ref{re1eqapp1_3}) into (\ref{PABand}b) yields
\begin{equation}\label{eqapp1_2}
B =\sum_{k=1}^K \max\left\{\frac {(\ln 2)R_{k}}{u^{-1}\left({\frac{\lambda} {F_{k}}}\right)}, {W_k} \right\}
\triangleq  \hat u(\lambda).
\end{equation}

Equation (\ref{eqapp1_2}) has a unique solution $\lambda>0$.
Since $u(x)$ is strictly increasing,
inverse function $u^{-1}(x)$ is also strictly increasing in $(0, +\infty)$.
Thus, $\hat u(\lambda_i)$ is a strictly decreasing function in $(0, +\infty)$.
Owing to the fact that $\lim_{\lambda \rightarrow 0+} \hat u(\lambda) =+\infty$
and $\lim_{\lambda \rightarrow +\infty} \hat u(\lambda) =0$,
there exists one unique $\lambda$ satisfying $\hat u(\lambda)=B$, and the solution can be obtained by using the bisection method.
Having obtained the value of $\lambda$, the optimal $\pmb w$ can be obtained from (\ref{re1eqapp1_3}).}}

\vspace{-1em}
\subsection{Iterative Algorithm and Complexity Analysis}
\vspace{-1em}
\begin{algorithm}[h]
\caption{\!\!: Iterative Algorithm}
\begin{algorithmic}[1]
\STATE Set the initial solution $( H^{(0)}, \Theta^{(0)}, \pmb y^{(0)},\pmb w^{(0)})$, and iteration number $n=1$.
\REPEAT
\STATE With fixed $\pmb y^{(n-1)}$ and $\pmb w^{(n-1)}$, obtain the optimal $(H^{(n)}, \Theta^{(n)})$ of problem (\ref{PAOABmin1}).
\STATE With fixed $(H^{(n)},\Theta^{(n)})$ and $\pmb w^{(n-1)}$, obtain the optimal $\pmb y^{(n)}$ of problem (\ref{PALPmin1}).
\STATE With fixed $(H^{(n)}, \Theta^{(n)})$ and $\pmb y^{(n)}$, obtain the optimal $\pmb w^{(n)}$ of problem (\ref{PABand}).
\STATE Set $n=n+1$.
\UNTIL the objective function (\ref{PAmin2}a) converges.
\end{algorithmic}
\end{algorithm}
The iterative procedure for solving Problem (\ref{PAmin2}) is given in Algorithm 1.
The main complexity of Algorithm 1 lies in Problem (\ref{PAOABmin1}) and Problem (\ref{PABand}).
For Problem (\ref{PAOABmin1}), the major computation comes from case 2, which needs to solve Problem (\ref{PAOABmin3}) via a one-dimension exhaustive search method with complexity $\mathcal O(\frac{\Theta_{\max}-\Theta_{\min}}{\delta})$ for minimal step $\delta$.
To solve Problem (\ref{PABand}), the major complexity lies in solving (\ref{eqapp1_2}) with complexity $\mathcal O(\log_2(1/\epsilon))$ for the bisection method with accuracy $\epsilon$.
As a result, the total complexity of Algorithm 1 is $\mathcal O(L_{\text {it}}\frac{\Theta_{\max}-\Theta_{\min}}{\delta}+L_{\text{it}}
\log_2(1/\epsilon))$, where $L_{\text{it}}$ is the number of iterations of the iterative Algorithm 1.

\vspace{-1em}
\section{Numerical Results}
\vspace{-0.5em}
We consider that there are $K=20$ GTs uniformly distributed in a circular area with radius $300$ m. 
We set $g_0=1.42\times 10^{-4}$, $B=10$ MHz, {\color{myc1}{$P_1=\cdots=P_K=20$ dBm}},
$\sigma^2=-169$ dBm/Hz, 
 $H_{\min}=50$ m, $H_{\max}=500$ m, $\Theta_{\min}=0$,
and $\Theta_{\max}=\pi/2$ rad.
We consider equal minimal rate demand, i.e., $R_1=\cdots=R_K=R$.

In Fig.~1, we consider the sum power (\ref{PAOABmin3}a), {\color{myc1}{which equals to (\ref{PAmin2}a) with fixed $\pmb y$ and $\pmb w$}}, versus $\Theta$ for various minimal rate demands with equal bandwidth allocation and the UAV located at the center of the circle.
With given minimal rate demand,
it is observed that the sum power first decreases and then increases with the increase of $\Theta$, which verifies the theoretical analysis in Section III.A.
\begin{figure}
\centering
\includegraphics[width=2.0in]{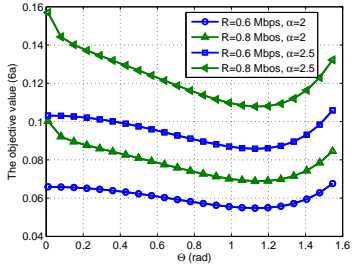}
\vspace{-1em}
\caption{Sum power versus $\Theta$.}
\vspace{-1em}
\end{figure}

We compare the proposed algorithm with the following four methods: fixed location method with optimized altitude, beamwidth and bandwidth (labeled as `FL'), fixed altitude and beamwidth method with optimized location and bandwidth (labeled as `FAB'), {\color{myc1}{fixed bandwidth method with optimized location altitude and beamwidth (labeled as `FB'), and exhaustive method via running Algorithm 1 with 1000 initial points (labeled as `Exhaustive')}}.
We investigate the sum power versus the minimal rate demand in Fig. 2.
It can be seen that the proposed algorithm outperforms FL, FAB and FB, especially when the minimal rate demand is large.
This is because the proposed algorithm jointly optimizes altitude, beamwidth, location and bandwidth.
{\color{myc1}{It can be seen that the sum power of the exhaustive method is slightly lower than that of the proposed algorithm, which indicates that the proposed algorithm approaches the globally optimal solution.}}

\begin{figure}
\centering
\includegraphics[width=2.0in]{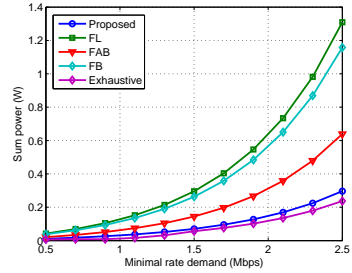}
\vspace{-1em}
\caption{Sum rate versus the minimal rate demand with $\alpha=2$. }
\vspace{-1em}
\end{figure}
\vspace{-1em}
\section{Conclusion}
\vspace{-0.5em}
In this letter, we {\color{myc1}{investigated}} the sum power minimization
problem in uplink UAV-enabled communications.
We {\color{myc1}{showed}} that the sum power first decreases and then decreases with the beamwidth.
Numerical results {\color{myc1}{showed}} that the uplink sum power performance can be improved by the proposed algorithm.
\vspace{-1.5em}
\bibliographystyle{IEEEtran}
\bibliography{IEEEabrv,MMM}

\begin{thebibliography}{10}
\providecommand{\url}[1]{#1}
\csname url@samestyle\endcsname
\providecommand{\newblock}{\relax}
\providecommand{\bibinfo}[2]{#2}
\providecommand{\BIBentrySTDinterwordspacing}{\spaceskip=0pt\relax}
\providecommand{\BIBentryALTinterwordstretchfactor}{4}
\providecommand{\BIBentryALTinterwordspacing}{\spaceskip=\fontdimen2\font plus
\BIBentryALTinterwordstretchfactor\fontdimen3\font minus
  \fontdimen4\font\relax}
\providecommand{\BIBforeignlanguage}[2]{{%
\expandafter\ifx\csname l@#1\endcsname\relax
\typeout{** WARNING: IEEEtran.bst: No hyphenation pattern has been}%
\typeout{** loaded for the language `#1'. Using the pattern for}%
\typeout{** the default language instead.}%
\else
\language=\csname l@#1\endcsname
\fi
#2}}
\providecommand{\BIBdecl}{\relax}
\BIBdecl

\bibitem{Zeng2017Mag}
Y.~Zeng, R.~Zhang, and T.~J. Lim, ``Wireless communications with unmanned
  aerial vehicles: {O}pportunities and challenges,'' \emph{IEEE Commun. Mag.},
  vol.~54, no.~5, pp. 36--42, May 2016.

\bibitem{6863654}
A.~Al-Hourani, S.~Kandeepan, and S.~Lardner, ``Optimal {LAP} altitude for
  maximum coverage,'' \emph{IEEE Wireless Commun. Lett.}, vol.~3, no.~6, pp.
  569--572, Dec. 2014.

\bibitem{7412759}
M.~Mozaffari, W.~Saad, M.~Bennis, and M.~Debbah, ``Unmanned aerial vehicle with
  underlaid device-to-device communications: {P}erformance and tradeoffs,''
  \emph{IEEE Trans. Wireless Commun.}, vol.~15, no.~6, pp. 3949--3963, Jun.
  2016.

\bibitem{7762053}
J.~Lyu, Y.~Zeng, R.~Zhang, and T.~J. Lim, ``Placement optimization of
  {UAV}-mounted mobile base stations,'' \emph{IEEE Commun. Lett.}, vol.~21,
  no.~3, pp. 604--607, Mar. 2017.

\bibitem{7888557}
Y.~Zeng and R.~Zhang, ``Energy-efficient {UAV} communication with trajectory
  optimization,'' \emph{IEEE Trans. Wireless Commun.}, vol.~16, no.~6, pp.
  3747--3760, Jun. 2017.

\bibitem{7875131}
M.~Chen, M.~Mozaffari, W.~Saad, C.~Yin, M.~Debbah, and C.~S. Hong, ``Caching in
  the sky: {P}roactive deployment of cache-enabled unmanned aerial vehicles for
  optimized quality-of-experience,'' \emph{IEEE J. Sel. Areas Commun.},
  vol.~35, no.~5, pp. 1046--1061, May 2017.

\bibitem{He2017CL}
H.~He, S.~Zhang, Y.~Zeng, and R.~Zhang, ``Joint altitude and beamwidth
  optimization for {UAV}-enabled multiuser communications,'' \emph{IEEE Commun.
  Lett.}, vol.~PP, no.~99, pp. 1--1, 2017.

\bibitem{constantine2016antenna}
A.~B. Constantine \emph{et~al.}, \emph{Antenna {T}heory: {A}nalysis and
  {D}esign}.\hskip 1em plus 0.5em minus 0.4em\relax 4th ed. New York: Wiley,
  2016.

\bibitem{boyd2004convex}
S.~Boyd and L.~Vandenberghe, \emph{Convex {O}ptimization}.\hskip 1em plus 0.5em
  minus 0.4em\relax Cambridge University Press, 2004.

\bibitem{Yang2017TVT}
Z.~Yang, C.~Pan, W.~Xu, H.~Xu, and M.~Chen, ``Joint time allocation and power
  control in multicell networks with load coupling: {E}nergy saving and rate
  improvement,'' \emph{IEEE Trans. Veh. Technol.}, vol.~66, no.~11, pp.
  10\,470--10\,485, Nov. 2017.

\end{thebibliography}

\end{document}